\documentclass[aps,prl,showpacs,twocolumn]{revtex4}
\usepackage{epsfig,color,times}

\begin{document}

\title{Stationary quantum statistics of a non-Markovian atom laser}
\author{A. S. Bradley$^{1,\dag}$, J. J. Hope$^{2}$, M. J.
Collett$^{3}$}
\affiliation{$^{1}$School of Chemical and Physical Sciences, Victoria
University of Wellington, New Zealand. \\
$^{2}$Department of Physics and Theoretical Physics, Australian
National University,
ACT 0200, Australia.\\
$^{3}$Department of Physics,
University of Auckland, New Zealand. \\
$^{\dag}$ email: ashton.bradley@vuw.ac.nz }
\date{\today}
\begin{abstract}
We present a steady state analysis of a quantum-mechanical model
of an atom laser. A single-mode atomic trap coupled to a continuum
of external modes is driven by a saturable pumping mechanism. In
the dilute flux regime, where atom-atom interactions are
negligible in the output, we have been able to solve this model
without making the Born-Markov approximation. The more exact
treatment has a different effective damping rate and occupation of
the lasing mode, as well as a shifted frequency and linewidth of
the output. We examine gravitational damping numerically, finding
linewidths and frequency shifts for a range of pumping rates. We
treat mean field damping analytically, finding a memory function
for the Thomas-Fermi regime. The occupation and linewidth are
found to have a nonlinear scaling behavior which has implications
for the stability of atom lasers.
\end{abstract}

\maketitle

\section{Introduction}
The development of Bose-Einstein condensation (BEC) \cite{BEC} has opened up the study of the atom laser: a coherent source of atoms analogous to the optical laser \cite{AL}.  These atomic sources may demonstrate many of the advantageous properties of optical lasers, such as spatial and temporal coherence, and high spectral flux.  This paper examines a quantum mechanical model of an atom laser consisting of a single-mode BEC coupled to a continuous output field, which may be a valid model for stable lasing systems when they have reached steady state.  We find a compact form for the standard quantum limit for such a laser model which can be determined without making the Born-Markov approximation.  This leads to nonlinear scaling of the steady-state occupation of the lasing mode with pumping rate, which has implications for the design of linewidth narrowing quantum feedback schemes and for the stability of atom lasers.

Outcoupling from a BEC has been achieved using radio frequency (rf) radiation \cite{Mewes97,Bloch99} and Raman transitions \cite{Hagley99} to change the internal state of the atoms to a non-trapped state.   Coupling the atoms out more slowly reduces the linewidth of the output at the expense of reducing the beam flux \cite{Hope97a}.  Optical lasers achieve a narrow linewidth through a competition between a saturable pumping mechanism and the damping of the lasing mode.  This allows a higher pumping rate to increase both the total flux and the spectral density of the output.  An atom laser with gain narrowing must also have a saturable pumping mechanism that operates at the same time as the damping \cite{Wiseman97}.  A recent experiment managed to combine a continuously produced series of condensates to maintain a BEC in a trap indefinitely \cite{Chikkatur02}.  This procedure was an excellent first step towards providing a continuous pumping mechanism.  The final step will be to produce a stimulated transition into the final BEC, which will allow mode-locking and reduce the phase diffusion of the lasing mode.  There have been several proposals for producing this kind of transition \cite{ctsPumpProp}, but they have not yet been experimentally achieved.

The analogy between optical lasers and atom lasers has some limits.  Theoretical examination of the properties of the output of atom lasers is complicated by the slow dispersion of atoms and the high interactions in atomic systems, which means that in general they exhibit multimode behaviour and the Born-Markov approximation cannot be made \cite{Moy99,Hope00,Haine02}.  Atom laser models have made either a single-mode approximation for the lasing state or the mean-field approximation \cite{BEC}.  The first class of models cannot examine the multimode behaviour of the laser, which has been shown to have significant effects on the stability of the device \cite{Haine02}.   The second class cannot examine the linewidth of an atom laser, which is a function of the quantum statistics of the lasing mode \cite{Walls}.  Many have also made the Born-Markov approximation, which has been shown to be invalid.  This paper examines the output properties of an atom laser which has reached steady state, which means that the quantum statistics of the lasing mode must be included in the model.  In order to avoid a multimode quantum field analysis, we assume that the steady state BEC can be modelled by a single mode, although we make no assumptions in our formalism about the spatial structure of that mode.

\section{The model}
\label{model} 
\begin{figure}
\epsfig{file=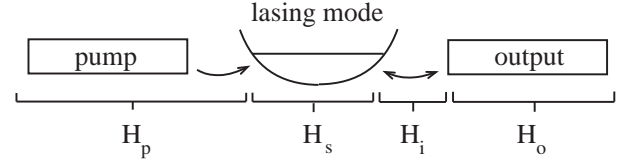,width=8cm}
\caption{\label{Fig1}Schematic of the atom laser model.}
\end{figure}

The atom laser is modelled by separating it into
three parts. Recent semiclassical modelling has shown that multimode
behaviour of the BEC can be very important, often determining the
stability of the system \cite{Haine02}.  Once conditions for stability have
 been achieved, by definition we expect the system to converge to a
stationary state.  While the details
of this stationary state will depend on the trapping conditions, pumping
method and damping rate, we will assume in this work that in the steady state the
BEC can be adequately described by a single mode.  This single-mode
BEC, also called the 'lasing' or 'system' mode, is modelled by the
annihilation(creation) operator
$a^{(\dag)}$ and Hamiltonian $H_{s}$. Since the external field and
trapped atoms are in a different electronic state, the free atoms are
not necessarily affected by the trapping potential. We describe
the continuum of external modes with field operators
$c_{p}^{(\dag)}$ and Hamiltonian $H_{o}$. The label $p$ is a
parameterization of the output eigenstates which in general may be
degenerate. The coupling between the lasing mode and the output
modes will be described by the interaction Hamiltonian $H_{i}$. At
this stage, we will describe the pump which irreversibly couples
the atoms from the pump reservoir into the trap mode by $H_{p}$.
The total Hamiltonian is then written
\begin{equation}
    H_{tot} = H_{p} + H_{s} + H_{i} + H_{o}
    \label{TotalH}
\end{equation}
where
\begin{eqnarray}
    H_{s} & = & \hbar \omega_{o} a^{\dag} a,
    \label{Hsys}  \\
    H_{i} & = & i\hbar \sqrt{\gamma} \int dp \; [\;\kappa^*(p,t)
        \;a^{\dag}c_{p}-\kappa(p,t)\;a\;c_{p}^{\dag}\;],
    \label{Hint} \\
    H_{o}&=&\int dp \;\hbar\omega_{p}\; c_{p}^{\dag}c_{p}\;,
\end{eqnarray}
and the operators satisfy
\begin{eqnarray}
\label{com1}
    [a,a^{\dag}] &=& 1,\\
    \label{com2}
    [c_{p},c^{\dag}_{p^{\prime}}] &=& \delta(p-p^{\prime})
\end{eqnarray}
with all other commutators zero. The natural
description of the output field in the dilute regime is in the
basis of energy eigenstates of the output Hamiltonian. The output
field eigenfunctions satisfy
\begin{equation}
H_{o}|u_{p}\rangle=\hbar\omega_{p}|u_{p}\rangle.
\end{equation}
In the position representation
the effective coupling function between the trapped mode and the
output field is the product of the condensate wave function and
the spatial profile of the coupling field. We denote the effective
outcoupled atomic state ket by $|\kappa,t\rangle = \int dp\;
\kappa(p,t)|u_{p}\rangle$.   We have taken out the overall coupling
rate $\sqrt{\gamma}$
in the interaction Hamiltonian so that $\kappa(p,t)$ satisfies $\int
dp\;|\kappa(p,t)|^{2}=1$.
The system is described in Fig.~\ref{Fig1}.

To carry out a stationary analysis we assume the coupling
amplitude is constant in time, apart from the Rabi frequency of
the coupling which may be eliminated by moving into a rotating
frame. This corresponds to a constant system wavefunction and
spatial profile of the coupling.

\subsection{Reversible output coupling}
Atomic output coupling has been achieved through Raman transitions
and radio frequency coupling of the trapped atoms to an un-trapped
or anti-trapped magnetic sublevel \cite{Mewes97,Bloch99,Hagley99}.
Later, we will consider the rf-coupling since in practice the energy shift
due to rf-coupling is negligible, unlike the Raman process which
imparts a significant impulse during coupling. Such an impulse is
modelled in our formalism by shifting the center of coupling function $\kappa$ in
momentum space. Any realistic model of an atom laser should describe
the coupling process in full as the usual Born and
Markov approximations which are so useful in quantum optics are
not generally applicable to atomic output couplers \cite{Moy99}.
In making the Markov approximation for the atomic output coupler,
one effectively assumes that the output process is irreversible,
so that outcoupled atoms have only a minor influence on the system
dynamics. While this works well in the optical case, the inertia
of atoms alters the coupling process in a crucial way. An atom may
be coupled into an untrapped internal state, but because it will
linger in the interaction region for some time, and it may evolve
for a short time in the output potential, and then undergo a
transition back into the trapped state. The slow dissipation of
atoms may cause the occupation of the system mode to increase.

We wish to solve the equations of motion of this model in the
stationary long time limit. To clarify matters we will first
reiterate some elementary results. The second-quantized output
field operator for the output field can be expanded over the
orthogonal set of modes
\begin{eqnarray}\label{outfield}
\hat{\psi}(x) &=& \int dp\;\langle x|u_p\rangle c_p.
\end{eqnarray}
Transforming into an interaction picture leaving only $H_{i}$ simplifies the description of the
output field. The interaction picture operators are
\begin{eqnarray}
&&\hat{\psi}_I(x,t)=e^{iH_o(t-t_o)/\hbar}\hat{\psi}(x,t_o)e^{-iH_o(t-t_o)/\hbar},\nonumber\\\\
&&a_I(t)=e^{i(H_s+H_p)(t-t_o)/\hbar}a(t_o)e^{-i(H_s+H_p)(t-t_o)/\hbar}.\nonumber\\
\end{eqnarray}
We set $t_o=0$ in the rest of this work. The Green's function
propagator for the output field is
\begin{eqnarray}
&&G(x,y;t) = [\hat{\psi}_I(x,t),\hat{\psi}^\dag_I(x,t)] = \langle
x|e^{-iH_ot/\hbar}|y\rangle,\nonumber\\
\end{eqnarray}
where
\begin{eqnarray}
\hat{\psi}_I(x,t) = \int dp\;\langle x|u_p\rangle \; c_p(0)e^{-i\omega_p
t}.
\end{eqnarray}

\subsubsection{Equations of motion without pumping}
The memory function for the output beam is defined by
\cite{Jack99a}
\begin{eqnarray}
F(x,t) &\equiv& \langle x|e^{-iH_ot/\hbar}|\kappa\rangle = \int
dy\;G(x,y;t)\kappa(y).
\end{eqnarray}
This may be interpreted as the output wavefunction, after evolving
for time $t$ in the output field. The Heisenberg equations of
motion for the output and trap operators are
\begin{eqnarray}\label{outeom}
{dc^\dag_p(t)\over dt}& = & i\omega_p\;
c^\dag_p(t)-\sqrt{\gamma}\kappa(p)^*a^\dag(t)\nonumber
\\\\\label{trapeom} {da^\dag(t)\over dt}& = &
-{i\over\hbar}[a^\dag(t),H_p]+i\omega_o
a^\dag(t)\nonumber\\&&+\sqrt{\gamma}\int dp\;\kappa(p)c^\dag_p(t).
\end{eqnarray}
Using the formal solution of (\ref{outeom}) with the Heisenberg
picture version of (\ref{outfield}), leads to the linear integral
equation for the output field
\begin{eqnarray}
&&\hat{\psi}(x,t) = \hat{\psi}_I(x,t) - \sqrt{\gamma}\int_0^t
dt^\prime\;a(t^\prime)F(x,t-t^\prime).\nonumber\\
\end{eqnarray}
This shows how the output field may be
constructed if the evolution of the trap operator is known. This simple
form for the output field arises because
we are considering the case of a dilute output beam.

The output coupling is reversible so we require a description of
the way atoms are coupled back into the trap. The overlap of the
output wavefunction with the trap mode describes the probability
of this occuring, and is defined by \cite{Hope00}
\begin{eqnarray}\label{fdef}
f(t)&\equiv&\langle \kappa|e^{-iH_ot/\hbar}\rangle \kappa|\\\nonumber\\
&=&\int dx\;\int dy\;G(x,y;t)\kappa^*(x)\kappa(y).
\end{eqnarray}
The Greens function for the output potential of $H_o$ and the form
of the coupling completely describes the influence of the output
on the system evolution. In the limit where $f(t)$ becomes a delta
function, and the resulting equations of motion will be
effectively memoryless or Markovian. This corresponds to the
quantum optical limit of the coupling where photon loss from an
optical cavity may be well described by a very narrow memory
function \cite{Gardiner}.

\subsection{Continuous pumping}

The biggest hurdle in the development of a mode-locked, gain-narrowed
atom laser is the production of a continuous pumping mechanism which
can repopulate the lasing mode by a stimulated transition.  The best
experiments at this stage have managed to merge independently
produced condensates in a pulsed, but quasi-continuous manner, which
provides an excellent starting point but does not involve the stimulated
transition \cite{Chikkatur02}.  The stimulated
nature of the transition is necessary for two reasons.  The phase of the
lasing mode must be preserved, and the small linewidth of a laser is a
product of the competition between a damping process and a saturable,
Bose-enhanced pumping \cite{Wiseman97}.  It is a highly non-trivial
experimental problem to design and build such a pumping process, and
rather than address these issues, we will use a model process that should
exhibit the qualitatively similar features to any successful continuous
pumping process.
A pumping process which can satisfy these requirements can be
found in a model where cooled atoms in an excited state are passed
over the trap containing the lasing mode \cite{Hope97a}. The
photon emission of the atoms would be stimulated by the presence
of the highly occupied ground state, and they will make a
transition into that state and emit a photon \cite{Hope96c}. For a
sufficiently optically thin sample, which can be made possible by
having a very tight, effectively low dimensional trap, the photon
is unlikely to be reabsorbed, and the process is effectively
irreversible.

If the pump reservoir is sufficiently
isolated from the cavity and external fields, and the pumping
process is designed to be irreversible, then we may trace over the
pump reservoir states to produce a master equation term for the
reduced density matrix which describes only the cavity and
external fields.

We choose to model an optical cooling process rather than the more
experimentally successful evaporative cooling process as we are
particularly interested in designing a continuously pumped system
with a steady state. The reader is referred to the literature for
the details of such proposals \cite{Hope00,Hope97a,Hope96c}. After
we have traced over the pump modes, a term due to the effect of
the pump in the equations of motion for $\langle
a^{\dag}(t)\rangle$ $\langle a^{\dag}a\rangle(t)$ is obtained by
tracing over the trap modes in the number state basis
\cite{Hope00}. It is important to note that we are not tracing
over the output field modes, so the reduced density matrix spans
the output field as well as the trap field, The term due to the
pump in the master equation for the density matrix is
\begin{eqnarray}
(\dot{\rho})_{pump}=r{\cal D}[a^{\dag}](n_{s}+{\cal
A}[a^{\dag}])^{-1}\rho \label{eq:pumprho}
\end{eqnarray}
where $r$ is the saturated gain rate and $n_{s}$ is the saturation
boson number. The superoperators are
\begin{eqnarray}
{\cal A}[c]=\frac{1}{2}(c^{\dag}c\rho-\rho c^{\dag}c),\nonumber
\\
\nonumber
\\
{\cal D}[c]\rho=c\rho c^{\dag}-{\cal A}\rho.\nonumber
\end{eqnarray}
This is the same pumping term as derived by Wiseman \cite{Wiseman95a}.
In the trap number state basis we denote $\rho_{n,m}=\langle
n|\rho|m\rangle$, and the pump term becomes
\begin{eqnarray}
(\dot{\rho}_{n,m})_{pump} &=&
r\frac{\sqrt{nm}}{n_{s}+(n+m)/2}\rho_{n-1,m-1}\nonumber
\\
\nonumber\\
&&-r\frac{(n+m+2)/2}{n_{s}+(n+m+2)/2}\rho_{n,m}.
\end{eqnarray}
This leads to the term in the $\langle a^{\dag}a\rangle$ equation
\begin{eqnarray}\label{numpump}
&&\frac{d}{dt}\langle a^{\dag}a\rangle\Big{|}_{pump} =
r\sum_{n=0}^{\infty}\frac{n+1}{n+1+n_{s}}\rho_{n,n} \nonumber
\\
\nonumber
\\
&&=r\sum_{n=0}^{\infty}\left(\frac{n}{n+n_{s}}+\frac{n_{s}}{(n+n_{s})(n+1+n_{s})}\right)\rho_{n,n}\nonumber
\\
\nonumber
\\
&&=r\left(\frac{\bar{N}}{\bar{N}+n_{s}}+O(\frac{n_{s}}{\bar{N}^{2}})\right)
\end{eqnarray}
The last line is obtained by assuming that when the mean trapped
mode occupation $\bar{N}$ becomes large, we may treat $n$ as a
continuous variable and expand the brackets in a power series
about $\bar{N}$. In contributing to the linewidth it will be seen
that the correction is of order $r/\bar{N}^{3}$ since a further
division by $\bar{N}$ occurs in deriving the spectrum. Since we
consider the regime $n_{s}<\bar{N}$ and $1\ll\bar{N}$ we neglect
the correction hereafter. The equation of motion for $\langle
a^{\dag}\rangle$ can be treated similarly, but in addition it is
assumed that as the system is driven further above threshold gain
narrowing will occur, permitting the coherent state approximation
$\rho\simeq|\sqrt{\bar{N}}\rangle\langle \sqrt{\bar{N}}|$
\cite{Hope00}. The matrix elements become
\begin{eqnarray}
\rho_{n,m}\simeq{
e^{-\bar{N}}\bar{N}^{(n+m)/2}\over\sqrt{n!m!}}\rangle n|\langle m|
\end{eqnarray}
and the equation of motion for $\langle a^\dag(t)\rangle$ has the
pump contribution
\begin{eqnarray}\label{corrpump}
\frac{d}{dt}\langle
a^{\dag}(t)\rangle\Big{|}_{pump}&=&r\sum_{n=0}^{\infty}\frac{\sqrt{n}\rho_{n-1,n}}{2(n+n_{s})+1}\nonumber
\\
\nonumber
\\
&\simeq&\frac{r}{2(\bar{N}+n_{s})+1}\langle
a^{\dag}(t)\rangle \equiv P\langle a^{\dag}(t)\rangle.\nonumber\\
\end{eqnarray}
Expanding the exact expression in a power series about $1/\bar{N}$
and using the coherent state approximation shows the correction is
of order $\;r/\bar{N}^{3}$.

\subsubsection{Steady-state occupation in the Born approximation}
In order to easily compare our results with the familiar quantum
optical case it will be necessary to discuss the Born
approximation. The Born approximation assumes that the density
matrix $\rho$ can be written as a tensor product of the reduced
density matirx for system $\sigma$, and the reduced density matrix
for the the output field. It further assumes that the output field
remains in its original state. If we trace over the output field
modes, we obtain a master equation for the lasing mode which has
the same pumping term as given above. The reduced density now
describes only the cavity mode, and $\sigma_{n,n}=\langle
n|\sigma|m\rangle$ are c numbers. The damping term has the form
\begin{eqnarray}
(\dot{\sigma})_{\rm damp}&=&-\gamma\int_0^t du\;\{f(u)e^{i\omega_ot}[a^\dag a\sigma(t-u)\nonumber\\
&&-a\sigma(t-u)a^\dag]+H.c.\}.
\end{eqnarray}
Combining the pumping and damping terms we can find the equation of motion for the steady-state
trap occupation distribution, $p_n=\sigma_{n,n}$,
\begin{eqnarray}
\dot{p}_n(t)&=&{rn\over n+n_s}p_{n-1}(t)-{r(n+1)\over n+n_s+1}p_n(t)\nonumber\\
&&-2n\int_0^t du\; {\rm Re}(f(u)e^{i\omega_ot})p_{n}(t-u)\nonumber\\\nonumber\\
&&+2(n+1)\int_0^t du\;{\rm
Re}(f(u)e^{i\omega_ot})p_{n+1}(t-u).\nonumber\\
\end{eqnarray}
By setting the derivatives to zero and assuming that the functions $p_n(t)$ approach a constant
$p_n^{ss}$ in the long time limit, we find the recursion relation
\begin{eqnarray}
&&{rn\over n+n_s}p_{n-1}^{ss}+(n+1)2\gamma_{\scriptscriptstyle BM}p_{n+1}^{ss}\nonumber\\
&&-\left({r(n+1)\over n+n_s+1}-2n\gamma_{\scriptscriptstyle
BM}\right)p_{n}^{ss}=0,
\end{eqnarray}
where
\begin{eqnarray}
\gamma_{\scriptscriptstyle BM}=\gamma\int_0^\infty du\;{\rm Re}(f(u)e^{i\omega_ou}),
\end{eqnarray}
is the effective damping constant.  The solution
\begin{eqnarray}
p_n^{ss}={\cal N}{(r/2\gamma_{\scriptscriptstyle BM})^n\over (n+n_s)!}
\end{eqnarray}
is almost identical to that obtained for the optical laser
\cite{Walls}. The distribution is thermal for
$\;r/\gamma_{2\scriptscriptstyle BM}<n_s$, and for
$\;r/2\gamma_{\scriptscriptstyle BM}\gg n_s$, it approaches a
Poissonian distribution with mean $\bar{N}$ and variance $V$ given
by
\begin{eqnarray}
\bar{N}&=&{r\over 2\gamma_{\scriptscriptstyle BM}}-n_s,\\
V&=&\bar{N}+n_s.
\end{eqnarray}
\subsubsection{Frequency shift and linewidth in the Markov approximation}
If the evolution of the system mode is much slower than the
timescale of decay for the memory function $f(t)$, the Markov
approximation may be used to obtain a useful limit of the
equations of motion. Using the solution for $c_p(t)$, we may write
\begin{eqnarray}
{da^\dag(t)\over dt}&=&-{i\over\hbar}[a^\dag,H_p]+i\omega_o a^\dag(t)+\sqrt{\gamma}\xi^\dag(t)\nonumber\\
&&-\gamma\int_0^t du\;a^\dag(u)f^*(t-u),
\end{eqnarray}
where
\begin{eqnarray}
\xi(t)=\int dp\;\kappa(p)^*c_p(0)e^{-i\omega_p t}
\end{eqnarray}
is an operator-valued noise term generated by the initial state of
the output which serves to preserve the commutation relations. The
commutator is given by the memory function found above
\begin{eqnarray}
[\xi(t),\xi^\dag(t^\prime)]=f(t-t^\prime),
\end{eqnarray}
which decays on a timescale called the memory time \cite{Jack99a},
which we denote by $T_m$. This characterizes the timescale for
which an output atom may influence the system evolution. The
Markov approximation is made by moving to a rotating frame at the
system frequency $a(t)=\tilde{a}(t)e^{-i\omega_ot},
c_p(t)=\tilde{c}_p(t)e^{-i\omega_ot}$, in which, if $T_m$ is very
small compared with the timescale for evolution of the system
operator, we may take $\tilde{a}^\dag(t)$ out of the integrals and
obtain the equation of motion
\begin{eqnarray}
{d\tilde{a}^\dag(t)\over
dt}&=&-{i\over\hbar}[\tilde{a}^\dag,H_p]+i\Delta_{\scriptscriptstyle
M}\tilde{a}^\dag(t)
\\
&&-\gamma_{\scriptscriptstyle BM} \tilde{a}^\dag(t)+\sqrt{\gamma}\tilde{\xi}^\dag(t),
\end{eqnarray}
where
\begin{eqnarray}\label{DeltaM}
\Delta_{\scriptscriptstyle M}=\gamma\int_0^\infty du\;{\rm Im}(f(u)e^{i\omega_ou}).
\end{eqnarray}
We can now construct equations of motion for the occupation and
the first order correlation function ${\rm
g}^{(1)}(\tau)\equiv\lim_{t\rightarrow\infty}\langle
a^\dag(t+\tau)a(t)\rangle/\bar{N}$. In the Markov approximation
the equations of motion for the occupation and the two time
correlation are uncoupled, and the equation of motion for $\langle
a^\dag(t+\tau)a(t)\rangle$ is independent of $t$. The equations
are

\begin{eqnarray}
{d\bar{N}(t)\over dt}=
r\left({\bar{N}(t)\over\bar{N}(t)+n_s}\right)-2\gamma_{\scriptscriptstyle
BM}\bar{N}(t),
\end{eqnarray}
and
\begin{eqnarray}
{d\over dt}{\rm g}^{(1)}(t)=\left(P-\gamma_{\scriptscriptstyle
BM}+i(\omega_o+\Delta_{\scriptscriptstyle M})\right){\rm
g}^{(1)}(t).
\end{eqnarray}
 The steady state two-time correlation is an exponential leading to a Lorentzian spectrum. Using the
notation ${\rm g}^{(1)}(t)=\exp{\left(-\Gamma_{\scriptscriptstyle
M}|t|+i(\Delta_{\scriptscriptstyle M}+\omega_o)t\right)}$, the
steady state parameters are
\begin{eqnarray}\label{MarkovssN}
\bar{N}&=&{r\over 2\gamma_{\scriptscriptstyle
BM}}-n_s,\\\label{Markovssgamma} \Gamma_{\scriptscriptstyle
M}&=&{r\over2(\bar{N}+n_s)}-{r\over2(\bar{N}+n_s)+1}\nonumber\\
&=&{r\over4(\bar{N}+n_s)^2}+O\left(\bar{N}^{-3}\right),
\end{eqnarray}
and the frequency shift is given by (\ref{DeltaM}).
>From this point of view it is clear that the Born and Markov approximations are closely related if
we are only describing the trap; in
particular, the effective damping constants are identical. This is because the output
and trap spaces remain uncorrelated in both approximations.

\section{Equations of motion}
We are now in a position to find the non-Markovian equations of
motion including the effect of the pump. The equations of motion
for the system+output, (\ref{trapeom}) and (\ref{outeom}), are
linear and Markovian; however the equations of motion for the
system mode alone will depend on the output field dynamics through
the memory function $f(t)$. Using the pump terms from Equations
(\ref{numpump}) and (\ref{corrpump}), and substituting the formal
solution of (\ref{outeom}) into (\ref{trapeom}), the equations of
motion for the trapped mode are
\begin{widetext}
\begin{eqnarray}
\label{neom} {d \over dt}\langle a^\dag a\rangle(t) &=&
r{\bar{N}\over \bar{N}+n_s} -\gamma 2{\rm Re}\left(\int_0^t
du\;f(u)\langle
a^\dag(t)a(t-u)\rangle\right),\\\nonumber\\\label{feom}{\partial
\over\partial \tau}\langle a^\dag(t+\tau)a(t)\rangle &=&
(i\omega_o + P)\langle a^\dag(t+\tau)a(t)\rangle
-\gamma\int_0^{t+\tau}\;du f^*(t+\tau-u)\langle
a^\dag(u)a(t)\rangle,\\\nonumber
\end{eqnarray}
and for the output modes we find
\begin{eqnarray}\label{ceom}
{d\langle c_p^\dag c_p\rangle\over dt}&=&\gamma|\kappa(p)|^22{\rm
Re}\left(\int_0^t du\;e^{-i\omega_p(t-u)}\langle
a^\dag(t)a(u)\rangle\right).
\end{eqnarray}
\end{widetext}
 These equations of motion were first derived in
\cite{Hope00}, with a seemingly minor modification of the pump
term in (\ref{neom}). This will be discussed in the following
section as it turns out to have important implications for the
predictions of the model. The first two coupled
integro-differential equations describe the system and output
dynamics, starting from when the pump is initiated. We are
primarily interested in the stationary solutions, and once the
system reaches a steady state the output energy spectrum will not
directly be of interest as it is steadily growing. The long time
limit of (\ref{ceom}) is the spectrum of the output energy flux
which will become stationary. In this case we will see that the
relationship between the output flux spectrum and the power
spectrum of the trapped mode is in keeping with Fermi's golden
rule.

 The equations of motion (37) and (38) of \cite{Hope00} are identical to our equations (\ref{neom}) and (\ref{feom})
apart from the additional factor $\bar{N}/(\bar{N}+n_s)$ in
(\ref{neom}). The treatment of the pump term in \cite{Hope00},
where this factor was taken as unity, produced incorrect results
because a similar approximation was not made in the pump term in
(38) of \cite{Hope00}. We will discuss this inconsistency in more
detail once we have derived the stationary equations of motion.

\subsection{Stationary equations of motion}
The stationary solutions of (\ref{neom}) and (\ref{feom}) may be
found by breaking up the integrals and using the fact that the
memory function vanishes in the long time limit. We will assume
the system reaches a steady state at $t=t_{ss}$. The integral in
(\ref{neom}) may then be written as
\begin{widetext}
\begin{eqnarray}
\lim_{t\to\infty}\int_{0}^t du\;f(u)\;\langle
a^\dag(t)a(t-u)\rangle&=& \lim_{t\to\infty}\int_0^{t-t_{ss}}
du\;f(u)\;{\rm g}^{(1)}(u)\bar{N}+\lim_{t\to\infty}\int_{t-t_{ss}}^t du\;f(u)\;\langle a^\dag(t)a(t-u)\rangle \\\nonumber\\
&=&\int_0^{\infty} du\;f(u)\;{\rm g}^{(1)}(u)\bar{N}.
\end{eqnarray}
Similar manipulations for the integral of (\ref{feom}) lead to the
term
\begin{eqnarray}\label{feomint}
\lim_{t\to\infty}\int_{0}^{t+\tau} du\;f^*(t+\tau-u) \;\langle
a^\dag(t)a(t-u)\rangle &=&\int_{0}^{\infty}du\;f(u)^*\;{\rm
g}^{(1)}(\tau-u)\bar{N},
\end{eqnarray}
where we have assumed stationarity when $t$ is much larger than
the memory time $T_{m}$, defined as the support of the memory
function. Using these expression we find the stationary equations
of motion for the trapped mode as
\begin{eqnarray}\label{ssN}
r&=&\gamma(\bar{N}+n_s)2{\rm Re}\left(\int_0^\infty\;du\;f(u){\rm
g}^{(1)}(u)\right),\\\nonumber\\\label{ssg1} {d\over
d\tau}{\rm g}^{(1)}(\tau)&=&(i\omega_o+P){\rm g}^{(1)}(\tau)-\gamma\int_0^\infty\;du\;f^*(u){\rm g}^{(1)}(\tau-u),\\\nonumber\\
\label{ssflux} {d\langle c_p^\dag c_p\rangle\over
dt}&=&2\pi\gamma\bar{N}|\langle u_p|\kappa\rangle|^2{\rm
S}(\omega_p).
\end{eqnarray}
\end{widetext}
which are exact in the long time limit. Equations (\ref{ssN}) and
(\ref{ssg1}) may be solved self consistently for the mean
occupation and the two time correlation of the trap mode. The
variables are the pumping rate $r$, the saturation Boson number
$n_s$, the damping rate $\gamma$, the memory function $f(t)$, and
the system frequency $\omega_o$. One might suspect that the
equation for ${\rm g}^{(1)}(\tau{\rm g}^{(1)}($ would be amenable
to some kind of transform method; however, it is easily seen that
this approach does not lead to any major simplifications. If the
integral is decomposed into
$\left(\int_0^\tau+\int_\tau^\infty\right)d\tau$, the first term
is a convolution, but the second term is still important for small
$\tau$, and this is the region of interest. The present form has
the added simplicity that we only require the memory function
$f(t)$, and not its Laplace transform which is generally more
complicated.

In deriving the output flux expression we have defined the
stationary power spectrum of the trapped mode as
\begin{eqnarray}
{\rm S}(\omega)&\equiv&\lim_{t\to\infty}\int_{-t}^{t}d\tau\;{\langle a^{\dag}(t+\tau)a(t)\rangle\over2\pi\langle a^\dag a\rangle(t)}\;e^{-i\omega \tau}\\
&=&{1\over 2\pi}\int d\tau\;{\rm g}^{(1)}(\tau)\;e^{-i\omega
\tau}.
\end{eqnarray}
Equation (\ref{ssflux}) is then readily found from (\ref{ceom}) by
handling the integrals in a similar manner as for the trapped mode
equations. This expression is essentially Fermi's Golden rule for
the atom laser model. It is easily verified that this result implies
that our stationary results are exact within first order
perturbation theory \cite{Sakurai}. This is because the coupling
only has non-zero matrix elements between the trap and output
eigenspaces, so that all higher order perturbation expressions,
which arise from energy nonconserving virtual transitions between
output eigenstates, vanish identically.

If the spectrum for the trap is an exponential ${\rm
g}^{(1)}(\tau)=\exp{(-\bar{\Gamma} |\tau|+i\bar{\omega}\tau)}$,
corresponding to a Lorentzian spectrum, then in the limit that the
system linewidth becomes very narrow the flux spectrum becomes
\begin{eqnarray}\label{pfluxlim}
 {d\langle c_p^\dag c_p\rangle \over
dt}\to2\pi\gamma\bar{N}|\langle u_p|\kappa\rangle
|^2\delta(\omega_p-\bar{\omega}).
\end{eqnarray}
The amplitude of the output spectral flux is now determined by the
matrix elements $|\langle u_p|\kappa\rangle|^2$.
 Using (\ref{ssflux}) we may find the corresponding expression for
 the spatial correlation
\begin{eqnarray}\label{xflux}
&&{d\over
dt}\langle\psi^\dag(x^\prime)\psi(x)\rangle\nonumber\\
&&\;\;\;\;\;\;=2\pi\gamma\bar{N}\int
dp\;\phi_p^*(x^\prime)\phi_p(x)|\langle u_p|\kappa\rangle|^2{\rm
S}(\omega_p),
\end{eqnarray}
where $\phi_p(x)\equiv\langle x|u_p\rangle$ is the output
eigenfunction in the position representation. In deriving these
results the interactions in the output beam and the trap have been
neglected; however, because the equation for $\dot{c}_p(t)$
remains linear in $a(t)$ when the interactions for the trapped
mode are included, Equations (\ref{ssflux}) and (\ref{xflux}) for
the output flux also hold when interactions are significant for
the trapped mode. The beam may be considered a true atom laser
beam when the spatial coherence length given by (\ref{xflux}) is
significantly longer than the thermal de-Broglie wavelength of the
atoms \cite{Wiseman97}. It is clear from (\ref{xflux}) that this
will place restrictions on the spectral width of the trap and the
force that acts on the output atoms. As the spectrum approaches a
delta function in frequency, the coherence length will be
determined by the eigenfunctions that have the same energy as the
trapped mode.

Since we are interested in a highly occupied trapped mode our
description of the trap should, in the single mode model, include
a self-interaction term of the form $H_{\rm coll}=\hbar Ca^\dag
a^\dag a a$ in the Hamiltonian, where $C$ is the nonlinear
interaction strength. In the limit where the interactions
dominate, the spectrum becomes
\begin{eqnarray}
{\rm g}^{(1)}(\tau)=\exp[-\bar{N}(1-e^{2iC\tau})].
\end{eqnarray}
which has the interesting feature of periodic revivals at
$\tau=m\pi/C$, for integer $m$ \cite{Thomsen2002}. In this regime
the non-Markovian term in (\ref{feom}) will become unimportant.
However, there are regimes of interest where the interactions are
weak, corresponding to a very weak harmonic confinement, or to
relatively low occupation number. In this regime, the Markovian
effects will alter the effective occupation, producing significant
linewidth and frequency changes in the spectrum.

\subsection{The Self-Consistent Markov Approximation}
Although the traditional Born-Markov approximation cannot be made
for atom lasers, as the linewidth becomes narrow there is still a
timescale separation between the memory time of the outcoupling
process and the temporal coherence of the lasing mode.  This
allows us to deal with the effects of the system memory using the
self-consistent Markov approximation, which was first presented in
\cite{Hope00}. This involves assuming the spectrum is Lorentzian
and solving for the linewidth and spectral shift self consistently
in an appropriate rotating frame. We are now in position to find
the sufficient condition for the validity of this approximation in
the steady state. Since the form of ${\rm g}^{(1)}(t)$ in our
equations of motion is still unknown, we must make some assumption
about its form, if we are to progress further. However, since we
are interested in the regime where linewidth narrowing is expected
to occur, we will assume that the memory function $f(t)$ has a
short decay time compared with the decay of ${\rm g}^{(1)}(t)$. In
this regime the decay of ${\rm g}^{(1)}(t)$ may be neglected in
the integrals of (\ref{ssN}) and (\ref{ssg1}). This corresponds to
the case where the atoms are ejected from the trap rapidly
compared with the system coherence time. Making the ansatz ${\rm
g}^{(1)}(t)=\exp{(-\Gamma_{\scriptscriptstyle
SM}|t|)}\exp{(i(\omega_o+\Delta_{\scriptscriptstyle SM})t)}$,
where $\Gamma_{\scriptscriptstyle SM}$ and
$\Delta_{\scriptscriptstyle SM}$ are real constants, the equations
of motion become
\begin{eqnarray}\label{ssNM}
&&r=(\bar{N}+n_s)2\gamma\int_0^\infty du\;{\rm
Re}\left(f(u)e^{i(\omega_o+\Delta_{\scriptscriptstyle
SM})u}\right),\\\nonumber\\\label{ssg1NM}
&&i\Delta_{\scriptscriptstyle SM}-\Gamma_{\scriptscriptstyle
SM}=P\nonumber\\
&&-\gamma\int_0^\infty
du\;f^*(u)e^{-i(\omega_o+\Delta_{\scriptscriptstyle
SM})u}\;e^{\Gamma_{\scriptscriptstyle SM}(t-|t-u|)}.
\end{eqnarray}
As $\Gamma_{\scriptscriptstyle SM}^{-1}$ becomes smaller than
$T_m$ we can ignore the variation of the rightmost factor in the
second integral.  This approximation explicitly removes any time
dependence from the equation, as required for the consistence of
our ansatz. If we define the following pair of self-consistent
parameters that depend solely on the memory function:
\begin{eqnarray}
\label{gammaSM}
&&\gamma_{\scriptscriptstyle SM}=\gamma\int_0^\infty du\;{\rm Re}\left(f(u)e^{i(\omega_o+\Delta_{\scriptscriptstyle SM})u}\right)\\
\label{DeltaSM} &&\Delta_{\scriptscriptstyle
SM}=\gamma\int_0^\infty du\;{\rm
Im}\left(f(u)e^{i(\omega_o+\Delta_{\scriptscriptstyle
SM})u}\right),
\end{eqnarray}
then we can immediately determine the steady-state lasing mode occupation and linewidth of the atom laser:
\begin{eqnarray}
\label{NSM} \bar{N}&=&{r\over2\gamma_{\scriptscriptstyle
SM}}-n_s\\
\label{GammaSM} \Gamma_{\scriptscriptstyle
SM}&=&{r\over2(\bar{N}+n_s)}-{r\over2(\bar{N}+n_s)+1}\nonumber\\
&=&{r\over4(\bar{N}+n_s)^2}+O(\bar{N}^{-3}).
\end{eqnarray}
These equations now take a very similar form to the Markov
equations, but must be solved self consistently for the system
variables in the steady state. The possibility of a frequency
shift also arises when the more severe Markov approximation is
made, but this only depends on the bare trap frequency, and
corresponds to the limit $\Delta_{\scriptscriptstyle
SM}\ll\omega_o$ of (\ref{DeltaSM}). (Note that the linewidth,
while taking the same functional form as the usual Born-Markov
result, depends on $\bar{N}$ which is determined by (\ref{NSM}),
and hence will be changed when $\Delta_{\scriptscriptstyle SM
}\neq\Delta_{\scriptscriptstyle M }$).

Equations (\ref{DeltaSM}) and (\ref{GammaSM}) are our main results.
Comparison with the Markov expressions shows
that the frequency shift for the system mode and the effective
damping constant must now be found self consistently. If the
output damping is dominated by gravity and the wavefunction of the
system state is weakly dependent on $\bar{N}$ the memory function
will not depend on the system variables. If the mean field
interactions are dominant, the memory function will depend on the
system wavefunction, which depends on $\bar{N}$. This gives the
possibility of novel linewidth and frequency dependence on the
pumping.

\subsection{Level shifts and damping rates}
It is worth establishing the connection between our results and
the more familiar quantum optical theory.

In terms of the output eigenfunctions, the memory function $f(t)$
can be written as
\begin{eqnarray}
f(t)=\int dp\;|\langle\kappa|u_p\rangle|^2\;e^{-i\omega_p t}.
\end{eqnarray}
 We now evaluate the time integral over $f(t)$, and use
\begin{eqnarray}
{1\over z+i\epsilon}\to {\cal P}\left({1\over
z}\right)-i\pi\delta(z),
\end{eqnarray}
for $\epsilon\to 0$, where ${\cal P}$ denotes the principal value
part of the integral. We find
\begin{eqnarray}
&&\int_0^\infty du\;f(u)e^{i(\omega_o+\Delta_{\scriptscriptstyle
SM})u}\nonumber\\
&&\;\;=\pi\int
dp\;|\langle\kappa|u_p\rangle|^2\delta(\omega_p-\omega_o-\Delta_{\scriptscriptstyle
SM})\nonumber\\
&&\;\;\;\;-i{\cal P}\int dp\;{|\langle\kappa|u_p\rangle|^2\over
\omega_p-\omega_o-\Delta_{\scriptscriptstyle SM}}.
\end{eqnarray}
The damping and frequency shift take the form
\begin{eqnarray}\label{pertDelta}
\Delta_{\scriptscriptstyle SM}&=&\gamma{\cal P}\int
dp\;{|\langle\kappa|u_p\rangle|^2\over
\omega_o+\Delta_{\scriptscriptstyle
SM}-\omega_p},\\\label{pertgamma} \gamma_{\scriptscriptstyle
SM}&=&\gamma\pi\int
dp\;|\langle\kappa|u_p\rangle|^2\delta(\omega_o+\Delta_{\scriptscriptstyle
SM}-\omega_p).
\end{eqnarray}
The level shift now takes the familiar form of a principal value
integral modified by the fact that it is now an implicit equation
that must be solved
simultaneously with (\ref{pertgamma}). When
$\Delta_{\scriptscriptstyle SM}\ll \omega_o$, the shift may be
found by neglecting $\Delta_{\scriptscriptstyle SM}$ in the
denominator, and can be more easily evaluated using
(\ref{DeltaSM}) in the same approximation.

\section{Output coupling}
The potentials in the output field may be caused by mean field
interactions, gravity, or by a magnetic field. In this section we
discuss the memory functions for these process, which describe the
influence of the output beam on the trapped mode statistics. While
our model is only strictly valid in the noninteracting limit where
the trapped mode becomes the ground state of the harmonic
potential, we will also find the memory function for the case of a
more highly occupied Bose-Einstein condensate, because deriving a
simple description of this output coupling process for the
Thomas-Fermi regine is of general interest and some practical use.
\subsection{Gravity}
If the Greens function for the output potential is known, and the
position space coupling function $\langle x|\kappa,t\rangle$ is
simple enough to compute the necessary integrals, the memory
function may be found exactly. The memory functions for gravity
and the magnetically anti-trapped output state have been found
using the Gaussian form for the coupling \cite{Jack99a}
\begin{eqnarray}
\kappa(x)=\left({1\over\pi\sigma^2}\right)^{1/4}e^{-x^2/2\sigma^2}.
\end{eqnarray}
When the potential is $V(x)=-mgx$ the result is
\begin{eqnarray}
f(t)={\exp{\left\{-({\omega_gt\over2})^2-i({mg^2\over24\hbar})t^3\right\}}\over\sqrt{1+i\omega_kt}},
\end{eqnarray}
where the kinetic and potential energies assiciated with the
initial wavepacket are $\hbar\omega_k=\hbar^2/2m\sigma^2$, and
$\hbar\omega_g=mg\sigma$ respectively. The general effect of the
output coupling is to change the system frequency and the
effective damping. To show this for the case of gravitational
damping we use $\gamma = 2\times 10^5 s^{-2}$, and find the Markov
and self consistent solutions as the pumping is increased. We use
$n_s = 47$, and increase the pumping from $2\times 10^5\;{\rm
s}^{-1}$ to $2\times 10^6\;{\rm s}^{-1}$. We use an atomic mass of
$5\times 10^{-26}{\rm kg}$. The bare system frequency is taken as
$\omega_o = 2\pi \times 123\;{\rm Hz}$. We choose gravity as $g =
9.8\sin(0.1){\rm ms}^{-2}$. We will also take the width of the
coupling to be $\sigma=1.6\times 10^{-6}{\rm m}$, which
corresponds to the momentum width used in \cite{Hope00}. Figure
\ref{fig2} shows the self consistent steady state occupation and
linewdith calculated for these parameters. The Markov
approximation neglects the effective weakening of the damping that
is caused by a large number of virtual transitions during the
output coupling. The more accurate self consistent approach
predicts an increase in $\bar{N}$ over the Markov results. The
difference in linewidths is a consequence of this effective
widening of the system boundary, and arises principally from the
difference in $\bar{N}$.

\subsection{Comparison with previous work}
\begin{figure}
\begin{center}
\epsfig{file=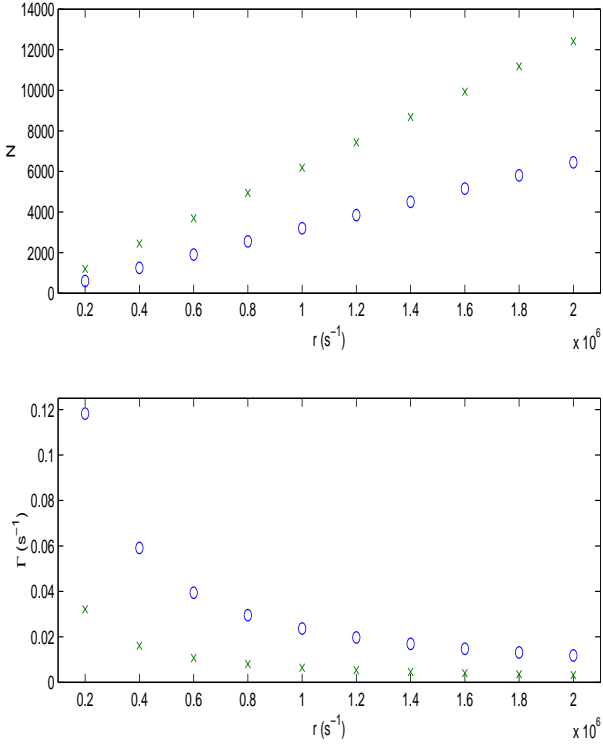,width=8cm,height=10cm}
\end{center}
\caption{Numerical results for the Markov approximation (o) and
the self consistent approach (x). We use $\gamma = 2\times 10^5
s^{-2}$, $n_s = 47$, $m=5\times 10^{-26}{\rm kg}$. The bare system
frequency is taken as $\omega_o = 2\pi \times 123\;{\rm Hz}$.
Gravity is set to $g = 9.8\sin(0.1){\rm ms}^{-2}$. The width of
the coupling is $\sigma=1.6\times 10^{-6}{\rm m}$. a) The steady
state occupation number as a function of $r$. b) The self
consistent linewidth as a function of $r$.} \label{fig2}
\end{figure}

In Figure \ref{fig2} we have used parameters corresponding to
Table I of \cite{Hope00}, which examined linewidth as a function
of pumping.  The previous treatment used the pump term $r$ instead of
$r\bar{N}/(\bar{N}+n_s)$ in (\ref{neom}). When stationary
solutions are found, instead of the self consistent Markov result
(\ref{GammaSM}), the linewidth becomes
\begin{eqnarray}
{r\over2\bar{N}}-{r\over2(\bar{N}+n_s)+1}={r (1+2 n_s)\over4\bar{N}^2}+O(\bar{N}^{-3}),
\end{eqnarray}
as can easily be verified using the data in Table I. of
\cite{Hope00}. This expression can also be written as
\begin{eqnarray}
{r\over2\bar{N}}-{r\over2(\bar{N}+n_s)+1}=\Gamma_{\scriptscriptstyle
SM}\left(1+2 n_s\right) + O(\bar{N}^{-3}),
\end{eqnarray}
which clearly generates a spurious broadening of the linewidth.
This masked the physical behaviour
found in the present work, where the linewidth is reduced below
the Born-Markov prediction by the non-Markovian evolution.


\subsection{Mean field interactions and antitrapping} When the
output atoms are coupled into an antitrapped state the memory
function should be calculated for an inverted parabolic potential.
If the atoms also experience significant mean field repulsion from
a highly occupied trap mode, the effective output potential will
be described by a rescaled inverted parabola corresponding to the
Thomas-Fermi solution for the condensate wavefunction. Using the
same Gaussian form of the coupling and taking the height of the
potential at the center of the trap as
$\tilde{V}=\hbar\tilde{\omega}$, the memory function is
\cite{Jack99a}
\begin{eqnarray}
f(t)=e^{-i\tilde{\omega}t}\left(\cosh{\omega_ot}+i\left({\omega_k\over\omega_o}-{\omega_o\over\omega_k}\right)\sinh{\omega_ot}\right)^{-1/2}
\end{eqnarray}
When $\omega_o^{-1}\ll t$, this becomes
\begin{eqnarray}
f(t)=e^{-i\tilde{\omega}t-\omega_ot/2}\left({1\over2}+{i\over2}\left({\omega_k\over\omega_o}-{\omega_o\over\omega_k}\right)\right)^{-1/2},
\end{eqnarray}
from which it is clear that the timescale of decay caused by the
kinetic dispersion associated with the curvature of the potential
is $2/\omega_o$. As noted in \cite{Jack99a}, this exponential
envelope is evident from the Greens function for the inverted
parabolic potential, and occurs regardless of the shape of the
interaction region. This is also expected to be the case for the
Thomas-Fermi mean field potential because while the exact motion
will not be correctly reconstructed by this potential outside the
interaction region, it is accurately reproduced where the overlap
is calculated.

Finding the output evolution for a particular potential is
equivalent to finding the Greens function for the output field
Schrodinger equation. This provides rather more information than
is needed, since we only require the evolution within the region
of overlap with the trap state. To find this, we will write the
output field effective Schrodinger equation in terms of amplitude
and phase variables. We may then use the Raman-Nath approximation
to evaluate the short time evolution when the trap is highly
occupied since the phase evolution generated by the mean field
will be dominant. We will also find the Raman-Nath time, and the
region of validity may then be found by comparing this with the
memory time arising from the approximate memory function. We will
first write the output wave function as $\Psi({\bf x},t)\equiv
A({\bf x},t)e^{iS({\bf x},t)}$, whereby the Schrodinger equation
reduces to
\begin{eqnarray}
{\partial A\over\partial t}&=&{-\hbar\over 2m}\left(2\nabla S\cdot\nabla
A+A\nabla^2S\right)\\\nonumber\\
{\partial S\over\partial t}&=&{\hbar\over 2m}\left({\nabla^2A\over
A}-(\nabla S)^2\right)-{V({\bf x})\over\hbar}.
\end{eqnarray}
To obtain the effective one dimensional equation of motion, the
radial wavefunction is assumed to be in the ground state of the
harmonic trapping potential, with frequency $\omega_r$. In terms
of the single particle condensate wavefunction $\psi_c$, the one
dimensional Schrodinger equation for the output atoms is
\begin{eqnarray}
i\hbar{\partial \psi\over\partial t}=-{\hbar^2\over 2m}{d^2\over dz^2}\psi+V(z)\psi,
\end{eqnarray}
where the effective potential is
\begin{eqnarray}
V(z)&=&\tilde{u}_{12}N|\psi_c|^2=\epsilon \mu_{\scriptscriptstyle TF}\left(1-{z^2\over L^2}\right)\\
&\equiv& V_c\left(1-{z^2\over L^2}\right),
\end{eqnarray}
where this is positive, and zero otherwise. The rescaled
interaction strength is $\tilde{u}_{12}=u_{12}/2\pi R^2$, and the
width of the ground state is determined by the harmonic oscillator
radius $R=\hbar/m\omega_r$. The axial Thomas-Fermi radius is
$L=\sqrt{2\mu_{\scriptscriptstyle TF}/m\omega_o^2}$ and the ratio
of trapped-trapped to trapped-untrapped s-wave scattering lengths
is $\epsilon$, which we take as unity in what follows. The
effective interaction parameter is rescaled by the transverse
cross section, and has dimension length$\times$energy as required
for a one dimensional model. The chemical potential becomes
\begin{eqnarray}
\mu_{\scriptscriptstyle
TF}=\left({3\bar{N}\tilde{u}_{11}\over4}\sqrt{m\omega_o^2\over2}\right)^{2/3}.
\end{eqnarray}

Starting from the density-phase equations for the output evolution
and reducing to one dimension, we neglect the curvature term
$\nabla^2A/A$, and take the initial phase as constant in position.
The phase equation then depends only on the potential, and may be
integrated to derive the memory function
\begin{eqnarray}\label{TFMapprx}
f_{\scriptscriptstyle TF}(t)={3\over
4}\sqrt{i\pi\over\omega_ct}e^{-i\omega_ct}h(\omega_ct),
\end{eqnarray}
where
\begin{eqnarray}
h(x)={e^{ix}\over\sqrt{-\pi ix}}+\left({1+2ix\over 2ix}\right){\rm
erf}\left(\sqrt{-ix}\right),
\end{eqnarray}
and $\omega_c=V_c/\hbar$. The long time behavior for $1\ll x$,
$h(x)\to 1$ has been used to factor out the asymptotic form of
$f(t)$.

The Raman-Nath time for the validity of this approximation is the
timescale of variation of the amplitude $A$. Using only the
potential for the phase evolution, and again using the smoothly
varying envelope approximation so that we drop the $\nabla
S\cdot\nabla A$ term, we find the short time solution
\begin{eqnarray}
A(z,t)=A(z,0)\exp{\left(-\omega_c{\hbar\over2mL^2}t^2\right)}.
\end{eqnarray}
Defining the kinetic energy associated with the length scale of
the Thomas-Fermi profile as $\hbar\omega_k=\hbar^2/2mL^2$, the
Raman-Nath time is
\begin{eqnarray}
\tau_{\scriptscriptstyle
RN}={1\over\sqrt{\omega_c\omega_k}}={2\over\omega_o}.
\end{eqnarray}
This is just the decay caused by the kinetic energy arising from
the curvature of the potential, as seen in our previous discussion
for the anti-trapped Gaussian wavepacket. Note that this does not
depend on $\bar{N}$ because the Thomas-Fermi solution always
assumes the form of an inverted image of the trap with amplitude
given by the chemical potential. We can find the condition for the
validity of the approximate memory function for our purposes by
comparing this timescale with the memory time. The memory time may
be defined by the ratio \cite{Jack99a}
\begin{eqnarray}
R={|\int_{T_m}^\infty dt f(t)|\over|\int_0^\infty dt f(t)|}.
\end{eqnarray}
When $T_m$ is chosen so that the ratio $R$ is small, the
contribution of any time integral equation involving $f(t)$ will
be significant for times less than $T_m$. We want to find $T_m$
when $R\ll 1$ for the Thomas-Fermi memory function
(\ref{TFMapprx}). We have
\begin{eqnarray}
R={\sqrt{\pi}\over2}\left|{e^{-i\omega_cT_m}{\rm erf}(\sqrt{-i\omega_cT_m})\over \sqrt{-i\omega_c
T_m}}\right|.
\end{eqnarray}
We now assume $1\ll \omega_cT_m$, and use the asymptotic form of
the memory function \cite{Abram} to find
\begin{eqnarray}
{\pi\over4\omega_cR^2}\lesssim T_m.
\end{eqnarray}
Using $R=10^{-2}$, the condition $T_m\ll\tau_{\scriptscriptstyle
RN} $ becomes
\begin{eqnarray}
\omega_o\ll 10^{-3}\omega_c.
\end{eqnarray}
When this condition holds the approximate memory function will
determine the effect of the output coupling on the system
evolution.
\subsubsection{Kinetic evolution}
We are primarily interested in the regime where the phase
evolution determines the memory function. However, if we include
the long-time exponential decay determined by the kinetic
evolution, the analysis is greatly simplified. This is because
neglecting the kinetic evolution entirely leads to an unphysical
long-time behavior of the memory function, which renders the
equations of motion insoluble. This feature is related to the
existence of a bound state in the dressed states of the trap plus
output potentials, which has been discussed in detail elsewhere
\cite{Hope00}. Since there is a separation of timescales, we may
include the kinetic evolution by simply multiplying the
approximate memory function by $e^{-\omega_o t/2}$. This has no
effect on the overlap $f(t)$ over the region in which the
integrand is significant, but restores the description of the long
time behavior to a more realistic form. The memory function that
we will use takes the final form
\begin{eqnarray}
f_{\scriptscriptstyle TF}(t)={3\over
4}\sqrt{i\pi\over\omega_ct}h(\omega_ct)e^{-i\omega_ct-\omega_o
t/2}.
\end{eqnarray}

\subsubsection{Mean field damping in the Markov approximation}
We will now find the linewidth and the frequency shift of the atom
laser model using $f_{\scriptscriptstyle TF}(t)$in the Markov
approximation, and show that mean field damping leads to an
interesting scaling behavior for the trap. We can compute an
analytic solution by using the integral
\begin{widetext}
\begin{eqnarray}
\int_0^\infty dt\;f_{\scriptscriptstyle
TF}(t)e^{-st}={3i\over2\omega_c}\left({i(s+\omega_o/2)\over
\sqrt{\omega_c(is+i\omega_o/2-\omega_c)}}\arctan{\sqrt{\omega_c\over
is+i\omega_o/2-\omega_c}}-1\right).
\end{eqnarray}\end{widetext}
This is easily found from the integral representation
\begin{eqnarray}
f_{\scriptscriptstyle TF}(t)={3\over 2}\int_0^1
dx\;(1-x^2)e^{-i\omega_c t(1-x^2)}\;e^{-\omega_o t/2},
\end{eqnarray}

which compactly expresses the approximations we are using for the
output evolution.

We now use the system frequency $\omega_o=\omega_c$ for the
condensate, and find that when $\omega_o\ll \omega_c$ the
asymptotic form of arctan \cite{Abram} leads to
\begin{eqnarray}
\Delta_{\scriptscriptstyle M}&=&\gamma_{\scriptscriptstyle
M}={3\pi\gamma\over 4\sqrt{\omega_c\omega_o}}.
\end{eqnarray}
The occupation becomes
\begin{eqnarray}
\bar{N}={2r\sqrt{\omega_c\omega_o}\over3\pi\gamma}-n_s,
\end{eqnarray}
so that with the $\bar{N}$ dependence of
$\omega_c=\mu_{\scriptscriptstyle TF}/\hbar$, well above threshold
the trap number becomes
\begin{eqnarray}\label{NTF}
\bar{N}= C\left({r\over \gamma}\right)^{3/2},
\end{eqnarray}
where
\begin{eqnarray}
C=\left({2\over3\pi}\sqrt{\omega_o\over \hbar
}\right)^{3/2}\left({3\tilde{u}_{11}\over 4}\sqrt{m\omega_o^2\over
2}\right)^{1/2}.
\end{eqnarray}
This is a scaling behaviour quite different to the optical case
which is varies as $\bar{N}\sim r/\gamma$. The trap spectrum
linewidth is
\begin{eqnarray}\label{GammaTF}
\Gamma_{\scriptscriptstyle M}={r\over 4\bar{N}^2}={\gamma^3\over
4(rC)^2}.
\end{eqnarray}
The possibility of the the damping depending on $\bar{N}$ leads to
some interesting new scaling behavior. The expression (\ref{NTF})
demonstrates that the Thomas-Fermi profile is not a particularly
good repelling potential, requiring that if the flux is increased,
the occupation must adjust by a correspondingly larger amount in
order to repel the extra atoms rapidly enough to recover
equilibrium. The linewidth scales as $\gamma^3/r^2$, rather than
the usual optical scaling of $\gamma^2/r$ which has consequences
for the response of atom laser statistics to fluctuations.

The validity of the Markov approximation is determined by the
condition $\Delta\ll \omega_c$, which leads to the requirement
\begin{eqnarray}
\left(\sqrt{2\over
m\omega_o^3}{\pi\gamma\over\tilde{u}_{11}}\right)^{2/3}\ll\bar{N}.
\end{eqnarray}
This forces a restriction on the rate of output coupling relative
to the interaction strength and the occupation.
\section{Conclusions}
We have carried out a non-Markovian steady state analysis of a
fully quantum mechanical atom laser model. We have demonstrated
that a self-consistent Markov approximation is valid provided the
laser is operating in a linewidth narrowing regime and the
reservoir correlation time is sufficiently short. We have shown
that the difference between the Born-Markov approximation and the
more exact treatment is that the system frequency, occupation
number and effective damping may be shifted by the coupling
process. This leads to corrections in the effective damping rate
and the steady state trap occupation number. We have found a
simple analytical form for the memory function for a Bose-Einstein
condensate output coupler in the Thomas-Fermi regime, which gives
rise to nonlinear scaling of the steady state occupation with
pumping rate. This has implications for the design of linewidth
narrowing quantum feedback schemes, and for the stability of atom
lasers.

\section{Acknowledgements}
This work was supported by the Marsden Fund of the Royal Society
of New Zealand. AB would like to thank M. K. Olsen for helpful and
stimulating comments.

\end{document}